\documentstyle[12pt]{article}

\begin{document}
\title{$\tau\rightarrow\rho\pi\pi\nu$ decays}
\author{Bing An Li\\
Department of Physics and Astronomy, University of Kentucky\\
Lexington, KY 40506, USA}

\maketitle
 
\begin{abstract}
Effective chiral theory of mesons is applied to study the four decay
modes of $\tau\rightarrow\rho\pi\pi\nu$. Theoretical values of the
branching ratios are in agreement with the data. The theory predicts
that the $a_{1}$ resonance plays a dominant role in these decays.
There is no new parameter in this study.
\end{abstract}

\newpage
Effective chiral theory of pseudoscalar, vector, and axial-vector
mesons proposed in Ref.[1] provides a unified description of meson
physics at low energies. In Ref.[2] it shows that the chiral
perturbation theory[3] is the low energy limit($<m_{\rho}$) of this
effective theory and the 10 coefficients of the chiral perturbation
theory have been predicted[2].
 
$\tau$ mesonic decays have been systematically studied by this
effective theory[4] and theoretical results agree with the data.
In this paper the decay modes of $\tau\rightarrow\rho\pi\pi\nu$
are investigated.
 
The branching ratios of $\tau\rightarrow4\pi\nu$ have been measured[5].
These decays are via CVC related to $ee^{+}\rightarrow4\pi$[6].
In Ref.[7] a dynamical model has been developed to study $4\pi$ modes of
$\tau$ decay. It is pointed out[7]
that the $a_{1}$ resonance is
important in understanding the data of $\tau\rightarrow4\pi\nu$.
 
This paper is the continuation of the study of $\tau$ mesonic decays
presented in Ref.[4]. The decays $\tau\rightarrow4\pi\nu$ are caused by
the vector current which is expressed as[4]
\begin{equation}
{\cal L}^{V}=\frac{g_{W}}{4}cos\theta_{C}g\{
-{1\over2}(\partial_{\mu}A^{i}_{\nu}-\partial_{\nu}A^{i}_{\mu})(
\partial^{\mu}\rho^{i\nu}-\partial^{\nu}\rho^{i\mu})	
+A^{i}_{\mu}j^{i\mu}\},
\end{equation}	
where $A^{i}_{\mu}$ is the W boson field, g is the universal coupling
constant defined in Ref.[1], and 
$j^{i}_{\mu}$ is the vector current	
derived by the substitution	
\begin{equation}
\rho^{i}_{\mu}\rightarrow\frac{g_{W}}{4}gcos\theta_{C}A^{i}_{\mu}	
\end{equation}
in the vertices involving $\rho$ meson.
 
All the vertices contributing to these decay modes
are derived from the Eq.(13) of Ref.[1a]. There are three kinds of
vertices. The first kind of vertices consists of three meson fields:
\begin{eqnarray}
\lefteqn{{\cal L}^{a_{1}\rho\pi}=\epsilon_{ijk}\{A(q^{2},p^{2})a^{i}_{\mu}
\rho^{j\mu}\pi^{k}-Ba^{i}_{\mu}\rho^{j}_{\nu}\partial^{\mu\nu}\pi^{k}
+Da^{i}_{\mu}\partial^{\mu}(\rho^{j}_{\nu}
\partial^{\nu}\pi^{k})\}},\\
&&A(q^{2},p^{2})={2\over f_{\pi}}f_{a}\{{F^{2}\over g^{2}}
+p^{2}[{2c\over g}+{3\over4
\pi^{2}g^{2}}(1-{2c\over g})]\nonumber \\
&&+q^{2}[{1\over 2\pi^{2}g^{2}}-
{2c\over g}-{3\over4\pi^{2}g^{2}}(1-{2c\over g})]\},\\
&&c={f^{2}_{\pi}\over2gm^{2}_{\rho}},\\
&&F^{2}=(1-{2c\over g})^{-2}f^{2}_{\pi},\\
&&f_{a}=(1-{1\over2\pi^{2}g^{2}})^{-{1\over2}},\\
&&B=-{2\over f_{\pi}}f_{a}{1\over2\pi^{2}g^{2}}(1-{2c\over g}),\\
&&D=-{2\over f_{\pi}}f_{a}\{{2c\over g}+{3\over2\pi^{2}g^{2}
}(1-{2c\over g})\},\\
&&{\cal L}^{\rho\pi\pi}={2\over g}\epsilon_{ijk}\rho^{i}_{\mu}
\pi^{j}\partial^{\mu}\pi^{k}-{2\over \pi^{2}f^{2}_{\pi}g}
\{(1-{2c\over g})^{2}-4\pi^{2}c^{2}\}\epsilon
_{ijk}\rho^{i}_{\mu}\partial_{\nu}\pi^{j}\partial^{\mu\nu}\pi^{k}
\nonumber \\
&&-{1\over \pi^{2}f^{2}_{\pi}g}\{3(1-{2c\over g})^{2}
+1-{2c\over g}-8\pi^{2}c^{2}\}\epsilon_{ijk}\rho^{i}_{\mu}\pi_{j}
\partial^{2}\partial_{\mu}\pi_{k},\\
&&{\cal L}^{\rho\rho\rho}=-{2\over g}\epsilon_{ijk}\partial_{\mu}
\rho^{i}_{\nu}\rho^{j\mu}\rho^{k\nu},\\
&&{\cal L}^{\omega\rho\pi}=-\frac{3}{\pi^{2}g^{2}f_{\pi}}
\varepsilon^{\mu\nu\alpha\beta}\partial_{\mu}
\omega_{\nu}\rho^{i}_{\alpha}\partial_{\beta}\pi^{i},
\end{eqnarray}
where q is the momentum of $a_{1}$ meson and p is the momentum of $\rho$
meson.
The second kind 
of vertices is contact interaction between two meson fields
\begin{eqnarray}
\lefteqn{{\cal L}^{\pi\pi}=\frac{1}{4\pi^{2}f^{2}_{\pi}}
(1-{2c\over g})^{2}\partial_{\mu\nu}\pi^{i}
\partial^{\mu\nu}\pi^{i}},\\
&&{\cal L}^{\pi a}=\frac{f_{a}}{2\pi^{2}gf_{\pi}}
(1-{2c\over g})\partial_{\mu\nu}
\pi^{i}\partial^{\mu}a^{i\nu},\\
&&{\cal L}^{aa}=\frac{f^{2}_{a}}{4\pi^{2}g^{2}}(\partial_{\mu}a^{i\mu})^{2}.
\end{eqnarray}
 
The third kind of vertices is the direct interaction between $\pi\pi\rho\rho$
\begin{eqnarray}
\lefteqn{{\cal L}^{\pi\pi\rho\rho}={4\over f_{\pi}}\epsilon_{ijk}
\epsilon_{i
j'k'}\{\frac{F^{2}}{2g^{2}}\rho^{j}_{\mu}\rho^{j'}_{\mu}
\pi_{k}\pi_{k'}
}\nonumber \\
&&+[-{2c^{2}\over g^{2}}+\frac{3}{4\pi^{2}g^{2}}(1-{2c\over g})^{2}]
\rho^{j}_{\mu}\rho^{j'}_{\nu}\partial^{\mu}\pi_{k}\partial^{\nu}\pi_{k'}
\nonumber \\
&&+[-{2c^{2}\over g^{2}}+\frac{1}{4\pi^{2}g^{2}}(1-{2c\over g})^{2}]
\rho^{i}_{\nu}\rho^{j'\nu}\partial_{\mu}\pi_{k}\partial^{\mu}\pi_{k'}
-{2c^{2}\over g^{2}}\rho^{j}_{\mu}\rho^{k'}_{\nu}\partial^{\nu}\pi_{k}
\partial^{\mu}\pi_{j'}\nonumber \\
&&-{3\over2\pi^{2}g^{2}}(1-{2c\over g})(\rho^{j}_{\mu}\pi_{k}\partial_{\nu}
\pi_{j'}-\rho^{j}_{\nu}\pi_{k}\partial_{\mu}\pi_{j'})\partial^{\nu}
\rho^{k'\mu}-{1\over2\pi^{2}g^{2}}(1-{2c\over g})\rho^{j}_{\mu}\pi_{k}
\rho^{j'}_{\nu}\partial^{\mu\nu}\pi_{k'}\nonumber \\
&&+{1\over4\pi^{2}g^{2}}[\partial_{\nu}(\rho^{j}_{\mu}\pi_{k})
\partial^{\nu}(\rho^{j}_{\mu}\pi_{k'})+2(1-{2c\over g})\rho^{j}_{\mu}\pi_{k}
\rho^{j'}_{\nu}\partial^{\mu\nu}\pi_{k'}]\}.
\end{eqnarray}

From these vertices(3-12) the physical processes contributing to $\tau
\rightarrow\rho\pi\pi\nu$ are obtained and shown in Fig.1. Similar diagrams 
involving the vertices(13-15) can be drawn too. 
The decay $\tau\rightarrow4\pi\nu$ originates in
$\tau\rightarrow\rho\pi\pi\nu$. 
 
In terms of Eq.(1) the vector current is derived from the vertices(3-16). 
In the
chiral limit this vector current must be conserved.
Derivative expansion is taken in the effective theory[1]. The vertices(3-16)
are derived up to the fourth order in derivatives.
Due to the anomaly the vector current conservation is guaranteed in Fig.(1e).	
It is proved that the vector current derived from Eqs.(3-11,13-15)
is indeed conserved in the
chiral limit.
The proof of vector current conservation shows that the presence
of the vertices(13-15)
is necessary to
make the vector current conserved. 	
 
There are four decay channels $\tau^{-}\rightarrow\rho^{0}
\pi^{-}\pi^{0}\nu,
\rho^{-}\pi^{+}\pi^{-}\nu, \rho^{-}\pi^{0}\pi^{0}\nu,
\rho^{+}\pi^{-}\pi^{-}\nu$.
 
The matrix element of the vector current
obtained from the diagram(fig.1(e))
is expressed as
\begin{eqnarray}
<\rho_{l}(p)\pi_{m}(p_{1})\pi_{n}(p_{2})|V^{-}_{\mu}|0>=
{1\over \sqrt{8E\omega_{1}\omega_{2}}}{g\over\sqrt{2}}(\delta_{i1}
-i\delta_{i2})\delta_{jl}\varepsilon_{\nu}^{ \mu\lambda\beta}
\varepsilon^{\nu\mu'\alpha\beta'}
\epsilon_{\alpha}q_{\lambda}p_{\mu'}({3\over \pi^{2}g^{2}f_{\pi}})^{2}
\nonumber \\
\frac{-m^{2}_{\rho}+i\sqrt{q^{2}}\Gamma_{\rho}(q^{2})}
{q^{2}-m^{2}_{\rho}+i\sqrt{q^{2}}\Gamma_{\rho}(q^{2})}
\{\delta_{im}\delta_{jn}\frac{p_{1\beta}p_{2\beta'}}{k^{2}_{1}
-m^{2}_{\omega}}+\delta_{in}\delta_{jm}\frac{p_{2\beta}p_{1\beta'}}
{k^{2}_{2}-m^{2}_{\omega}}\},
\end{eqnarray}
where \(k_{1}=q-p_{1}\) and \(k_{2}=q-p_{2}\).
 
The matrix element of the vector
current obtained from the diagrams(fig.1(a-d)) and the diagrams involving 
the vertices(13-15) is written as
\begin{eqnarray}
\lefteqn{<\rho_{l}(p)\pi_{m}(p_{1})\pi_{n}(p_{2})|V^{-}_{\mu}|0>=
{g\over\sqrt{8E\omega_{1}\omega_{2}}}{1\over\sqrt{2}}(\delta_{j1}
-i\delta_{j2})\delta_{j'l}\epsilon_{ijk}\epsilon_{ij'k'}
\epsilon_{\nu}\frac{-m^{2}_{\rho}+i\sqrt{q^{2}}\Gamma_{\rho}(q^{2})}
{q^{2}-m^{2}_{\rho}+i\sqrt{q^{2}}\Gamma_{\rho}(q^{2})}}\nonumber \\
&&\{\delta_{km}\delta_{k'n}[(g_{\mu\nu}-\frac{q_{\mu}q_{\nu}}{q^{2}})f
+(p_{1\mu}-\frac{p_{1}\cdot q}{q^{2}}q_{\mu})(f_{11}p_{1\nu}
+f_{12}p_{2\nu})
+(p_{2\mu}-\frac{p_{2}\cdot q}{q^{2}}q_{\mu})(f_{12}p_{1\nu}
+f_{22}p_{2\nu})]\nonumber \\
&&+\delta_{kn}\delta_{k'm}[(g_{\mu\nu}-\frac{q_{\mu}q_{\nu}}{q^{2}})f'
+(p_{2\mu}-\frac{p_{2}\cdot q}{q^{2}}q_{\mu})(f'_{11}p_{2\nu}
+f'_{12}p_{1\nu})
+(p_{1\mu}-\frac{p_{1}\cdot q}{q^{2}}q_{\mu})(f'_{12}p_{2\nu}
+f'_{22}p_{1\nu})]\},
\end{eqnarray}
where
\[q=p+p_{1}+p_{2},\]
where f' and $f'_{ij}$ are obtained from f and $f_{ij}$ by exchanging
$p_{1}\leftrightarrow p_{2}$, $\Gamma_{\rho}$ is the decay
width of $\rho$ meson of momentum q
\begin{equation}
\Gamma_{\rho}(q^{2})=\frac{q^{2}}{12\pi g^{2}m_{\rho}}\{1+
\frac{q^{2}}{2\pi^{2}f^{2}_{\pi}}[(1-{2c\over g})^{2}-
4\pi^{2}c^{2}]\}^{2}(1-{4m^{2}_{\pi}\over q^{2}})^{3\over2}.
\end{equation}
The contributions of
$a_{1}$ meson(fig.1b) to the functions f and $f_{ij}$ are presented below.
The contributions from other diagrams are shown in the appendix.
\begin{eqnarray}
\lefteqn{BW(k^{2}_{1})=\frac{1}{k^{2}_{1}-m^{2}_{a}
+i\sqrt{k^{2}_{1}}\Gamma_{a}
(k^{2}_{1})}},\nonumber \\
&&f=BW(k^{2}_{1})
A(q^{2},k^{2}_{1})A(m^{2}_{\rho},k^{2}_{1}),\nonumber \\
&&f_{11}=BW(k^{2}_{1})
A(m^{2}_{\rho},k^{2}_{1})B,\nonumber \\
&&f_{12}=BW(k^{2}_{1})
\{[A(m^{2}_{\rho},k^{2}_{1})+A(q^{2},k^{2}_{1})]
D+(k_{1}\cdot p_{2}-k_{1}\cdot p_{1})BD+p_{1}\cdot p_{2}B^{2}
-k^{2}_{1}D^{2}\}\nonumber \\
&&-BW(k^{2}_{1})
{1\over m^{2}_{a}}\{-A(q^{2},k^{2}_{1})+k_{1}\cdot
p_{1}B+k^{2}_{1}D\}\{A(m^{2}_{\rho},k^{2}_{1})+k_{1}\cdot p_{2}B
-k^{2}_{1}D\},\nonumber \\
&&f_{22}=BW(k^{2}_{1})A(q^{2},k^{2}_{1})B,
\end{eqnarray}
The decay width
of $a_{1}$ meson is derived as
\begin{eqnarray}
\Gamma_{a}(k^{2})={k_{a}\over 12\pi}{1\over \sqrt{k^{2}}m_{a}}
\{(3+{k^{2}_{a}\over m^{2}_{\rho}})A^{2}(m^{2}_{\rho},k^{2})
-2A(m^{2}_{\rho},k^{2})B(k^{2}+m^{2}_{\rho}){k^{2}_{a}
\over m^{2}_{\rho}}+{k^{2}\over m^{2}_{\rho}}k^{4}_{a}B^{2}\}.
\nonumber \\
k^{2}_{a}={1\over k^{2}}(k^{2}+m^{2}_{\rho}-m^{2}_{\pi})
-m^{2}_{\rho}.
\end{eqnarray}
 
In the matrix elements (17,18) there are three parameters: $f_{\pi}$,
$m_{\rho}$, and g. Obviously, the first two have been fixed.
and the third one is determined from $\rho\rightarrow ee^{+}$ to be \(g=0.39\).
The mass of the $a_{1}$ meson has been derived as[1]
\[(1-{1\over2\pi^{2}g^{2}})m^{2}_{a}={F^{2}\over g^{2}}+m^{2}_{\rho}.\]
Therefore, this study is parameter free.
 
The branching ratios
of the four decay channels are calculated
\begin{eqnarray}
\lefteqn{B(\tau\rightarrow\rho^{0}\pi^{0}\pi^{-}\nu)
=B(\tau\rightarrow\rho^{-}\pi^{+}\pi^{-}\nu)=1.02\%}\nonumber \\
&&B(\tau\rightarrow\rho^{-}\pi^{0}\pi^{0}\nu)
=B(\tau\rightarrow\rho^{+}\pi^{-}\pi^{-}\nu)=1.27\%.
\end{eqnarray}
The experimental data[8] are
\[B(\tau\rightarrow h^{-}\rho\pi^{0}\nu)=(1.33\pm0.20)\%\;\;\;
B(\tau\rightarrow h^{-}\rho^{-}h^{+}\nu)=(1.15\pm0.23)\%,\]
\[B(\tau\rightarrow h^{-}\rho^{+}h^{-}\nu)=(4.4\pm2.2)\times10^{-3}\;\;\;
B(\tau\rightarrow h^{-}(\rho\pi)^{0}\nu)=(2.84\pm0.34)\%.\]
Except for the $h^{-}\rho^{+}h^{-}$ mode the theoretical results
are in agreement with the data.
 
It is necessary to point out that in order to have the
current conservation all the diagrams must be
taken into account. However, the calculations show that
the $a_{1}$ meson(fig1(b)) dominates the four decay channels.
If only the diagram(fig.1(b)) is kept in the matrix elements(16)
it is obtained that
\[B(\tau\rightarrow\rho^{-}\pi^{-}\pi^{+}\nu)=1.25\%.\]
The dominance of the $a_{1}$ meson in other three channels
is true too. There are two reasons for this dominance:
\begin{enumerate}
\item The mass of the $a_{1}$ meson is the energy range of the
decay $\tau\rightarrow\rho\pi\pi\nu$,
\item the strong coupling between $a_{1}$ and $\rho\pi$(3) which
leads to the wide decay width of the $a_{1}$ meson[4].
\end{enumerate}
 
The distribution functions,
$\frac{d\Gamma}{d q^{2}}$, are calculated and shown in fig.2 and fig.3.

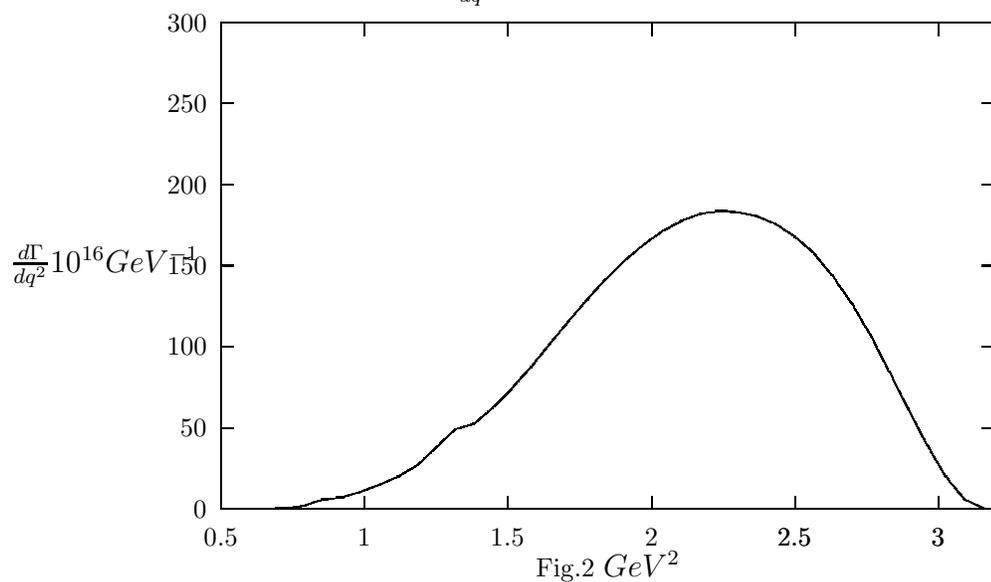
\setlength{\unitlength}{0.240900pt}
\ifx\plotpoint\undefined\newsavebox{\plotpoint}\fi
\sbox{\plotpoint}{\rule[-0.200pt]{0.400pt}{0.400pt}}%
\begin{picture}(1500,900)(0,0)
\font\gnuplot=cmr10 at 10pt
\gnuplot
\sbox{\plotpoint}{\rule[-0.200pt]{0.400pt}{0.400pt}}%
\put(220.0,113.0){\rule[-0.200pt]{292.934pt}{0.400pt}}
\put(220.0,113.0){\rule[-0.200pt]{4.818pt}{0.400pt}}
\put(198,113){\makebox(0,0)[r]{0}}
\put(1416.0,113.0){\rule[-0.200pt]{4.818pt}{0.400pt}}
\put(220.0,240.0){\rule[-0.200pt]{4.818pt}{0.400pt}}
\put(198,240){\makebox(0,0)[r]{50}}
\put(1416.0,240.0){\rule[-0.200pt]{4.818pt}{0.400pt}}
\put(220.0,368.0){\rule[-0.200pt]{4.818pt}{0.400pt}}
\put(198,368){\makebox(0,0)[r]{100}}
\put(1416.0,368.0){\rule[-0.200pt]{4.818pt}{0.400pt}}
\put(220.0,495.0){\rule[-0.200pt]{4.818pt}{0.400pt}}
\put(198,495){\makebox(0,0)[r]{150}}
\put(1416.0,495.0){\rule[-0.200pt]{4.818pt}{0.400pt}}
\put(220.0,622.0){\rule[-0.200pt]{4.818pt}{0.400pt}}
\put(198,622){\makebox(0,0)[r]{200}}
\put(1416.0,622.0){\rule[-0.200pt]{4.818pt}{0.400pt}}
\put(220.0,750.0){\rule[-0.200pt]{4.818pt}{0.400pt}}
\put(198,750){\makebox(0,0)[r]{250}}
\put(1416.0,750.0){\rule[-0.200pt]{4.818pt}{0.400pt}}
\put(220.0,877.0){\rule[-0.200pt]{4.818pt}{0.400pt}}
\put(198,877){\makebox(0,0)[r]{300}}
\put(1416.0,877.0){\rule[-0.200pt]{4.818pt}{0.400pt}}
\put(220.0,113.0){\rule[-0.200pt]{0.400pt}{4.818pt}}
\put(220,68){\makebox(0,0){0.5}}
\put(220.0,857.0){\rule[-0.200pt]{0.400pt}{4.818pt}}
\put(445.0,113.0){\rule[-0.200pt]{0.400pt}{4.818pt}}
\put(445,68){\makebox(0,0){1}}
\put(445.0,857.0){\rule[-0.200pt]{0.400pt}{4.818pt}}
\put(670.0,113.0){\rule[-0.200pt]{0.400pt}{4.818pt}}
\put(670,68){\makebox(0,0){1.5}}
\put(670.0,857.0){\rule[-0.200pt]{0.400pt}{4.818pt}}
\put(896.0,113.0){\rule[-0.200pt]{0.400pt}{4.818pt}}
\put(896,68){\makebox(0,0){2}}
\put(896.0,857.0){\rule[-0.200pt]{0.400pt}{4.818pt}}
\put(1121.0,113.0){\rule[-0.200pt]{0.400pt}{4.818pt}}
\put(1121,68){\makebox(0,0){2.5}}
\put(1121.0,857.0){\rule[-0.200pt]{0.400pt}{4.818pt}}
\put(1346.0,113.0){\rule[-0.200pt]{0.400pt}{4.818pt}}
\put(1346,68){\makebox(0,0){3}}
\put(1346.0,857.0){\rule[-0.200pt]{0.400pt}{4.818pt}}
\put(1121.0,113.0){\rule[-0.200pt]{0.400pt}{4.818pt}}
\put(1121,68){\makebox(0,0){2.5}}
\put(1121.0,857.0){\rule[-0.200pt]{0.400pt}{4.818pt}}
\put(1346.0,113.0){\rule[-0.200pt]{0.400pt}{4.818pt}}
\put(1346,68){\makebox(0,0){3}}
\put(1346.0,857.0){\rule[-0.200pt]{0.400pt}{4.818pt}}
\put(220.0,113.0){\rule[-0.200pt]{292.934pt}{0.400pt}}
\put(1436.0,113.0){\rule[-0.200pt]{0.400pt}{184.048pt}}
\put(220.0,877.0){\rule[-0.200pt]{292.934pt}{0.400pt}}
\put(45,495){\makebox(0,0){${d\Gamma\over dq^{2}}10^{16}GeV^{-1}$ }}
\put(828,23){\makebox(0,0){Fig.2   $GeV^{2}$}}
\put(220.0,113.0){\rule[-0.200pt]{0.400pt}{184.048pt}}
\put(262,113){\usebox{\plotpoint}}
\put(291,112.67){\rule{7.227pt}{0.400pt}}
\multiput(291.00,112.17)(15.000,1.000){2}{\rule{3.613pt}{0.400pt}}
\multiput(321.00,114.60)(4.283,0.468){5}{\rule{3.100pt}{0.113pt}}
\multiput(321.00,113.17)(23.566,4.000){2}{\rule{1.550pt}{0.400pt}}
\multiput(351.00,118.58)(1.486,0.491){17}{\rule{1.260pt}{0.118pt}}
\multiput(351.00,117.17)(26.385,10.000){2}{\rule{0.630pt}{0.400pt}}
\multiput(380.00,128.61)(6.490,0.447){3}{\rule{4.100pt}{0.108pt}}
\multiput(380.00,127.17)(21.490,3.000){2}{\rule{2.050pt}{0.400pt}}
\multiput(410.00,131.59)(1.718,0.489){15}{\rule{1.433pt}{0.118pt}}
\multiput(410.00,130.17)(27.025,9.000){2}{\rule{0.717pt}{0.400pt}}
\multiput(440.00,140.58)(1.345,0.492){19}{\rule{1.155pt}{0.118pt}}
\multiput(440.00,139.17)(26.604,11.000){2}{\rule{0.577pt}{0.400pt}}
\multiput(469.00,151.58)(1.171,0.493){23}{\rule{1.023pt}{0.119pt}}
\multiput(469.00,150.17)(27.877,13.000){2}{\rule{0.512pt}{0.400pt}}
\multiput(499.00,164.58)(0.838,0.495){33}{\rule{0.767pt}{0.119pt}}
\multiput(499.00,163.17)(28.409,18.000){2}{\rule{0.383pt}{0.400pt}}
\multiput(529.00,182.58)(0.526,0.499){109}{\rule{0.521pt}{0.120pt}}
\multiput(529.00,181.17)(57.918,56.000){2}{\rule{0.261pt}{0.400pt}}
\multiput(588.00,238.59)(1.718,0.489){15}{\rule{1.433pt}{0.118pt}}
\multiput(588.00,237.17)(27.025,9.000){2}{\rule{0.717pt}{0.400pt}}
\multiput(618.00,247.58)(0.580,0.497){47}{\rule{0.564pt}{0.120pt}}
\multiput(618.00,246.17)(27.829,25.000){2}{\rule{0.282pt}{0.400pt}}
\multiput(647.00,272.58)(0.499,0.497){57}{\rule{0.500pt}{0.120pt}}
\multiput(647.00,271.17)(28.962,30.000){2}{\rule{0.250pt}{0.400pt}}
\multiput(677.58,302.00)(0.497,0.549){57}{\rule{0.120pt}{0.540pt}}
\multiput(676.17,302.00)(30.000,31.879){2}{\rule{0.400pt}{0.270pt}}
\multiput(707.58,335.00)(0.497,0.621){55}{\rule{0.120pt}{0.597pt}}
\multiput(706.17,335.00)(29.000,34.762){2}{\rule{0.400pt}{0.298pt}}
\multiput(736.58,371.00)(0.497,0.600){57}{\rule{0.120pt}{0.580pt}}
\multiput(735.17,371.00)(30.000,34.796){2}{\rule{0.400pt}{0.290pt}}
\multiput(766.58,407.00)(0.497,0.583){57}{\rule{0.120pt}{0.567pt}}
\multiput(765.17,407.00)(30.000,33.824){2}{\rule{0.400pt}{0.283pt}}
\multiput(796.58,442.00)(0.497,0.551){55}{\rule{0.120pt}{0.541pt}}
\multiput(795.17,442.00)(29.000,30.876){2}{\rule{0.400pt}{0.271pt}}
\multiput(825.00,474.58)(0.499,0.497){57}{\rule{0.500pt}{0.120pt}}
\multiput(825.00,473.17)(28.962,30.000){2}{\rule{0.250pt}{0.400pt}}
\multiput(855.00,504.58)(0.600,0.497){47}{\rule{0.580pt}{0.120pt}}
\multiput(855.00,503.17)(28.796,25.000){2}{\rule{0.290pt}{0.400pt}}
\multiput(885.00,529.58)(0.692,0.496){39}{\rule{0.652pt}{0.119pt}}
\multiput(885.00,528.17)(27.646,21.000){2}{\rule{0.326pt}{0.400pt}}
\multiput(914.00,550.58)(0.945,0.494){29}{\rule{0.850pt}{0.119pt}}
\multiput(914.00,549.17)(28.236,16.000){2}{\rule{0.425pt}{0.400pt}}
\multiput(944.00,566.58)(1.392,0.492){19}{\rule{1.191pt}{0.118pt}}
\multiput(944.00,565.17)(27.528,11.000){2}{\rule{0.595pt}{0.400pt}}
\multiput(974.00,577.60)(4.137,0.468){5}{\rule{3.000pt}{0.113pt}}
\multiput(974.00,576.17)(22.773,4.000){2}{\rule{1.500pt}{0.400pt}}
\put(1003,579.17){\rule{6.100pt}{0.400pt}}
\multiput(1003.00,580.17)(17.339,-2.000){2}{\rule{3.050pt}{0.400pt}}
\multiput(1033.00,577.93)(2.247,-0.485){11}{\rule{1.814pt}{0.117pt}}
\multiput(1033.00,578.17)(26.234,-7.000){2}{\rule{0.907pt}{0.400pt}}
\multiput(1063.00,570.92)(1.229,-0.492){21}{\rule{1.067pt}{0.119pt}}
\multiput(1063.00,571.17)(26.786,-12.000){2}{\rule{0.533pt}{0.400pt}}
\multiput(1092.00,558.92)(0.753,-0.496){37}{\rule{0.700pt}{0.119pt}}
\multiput(1092.00,559.17)(28.547,-20.000){2}{\rule{0.350pt}{0.400pt}}
\multiput(1122.00,538.92)(0.576,-0.497){49}{\rule{0.562pt}{0.120pt}}
\multiput(1122.00,539.17)(28.834,-26.000){2}{\rule{0.281pt}{0.400pt}}
\multiput(1152.58,511.58)(0.497,-0.603){55}{\rule{0.120pt}{0.583pt}}
\multiput(1151.17,512.79)(29.000,-33.790){2}{\rule{0.400pt}{0.291pt}}
\multiput(1181.58,476.20)(0.497,-0.718){57}{\rule{0.120pt}{0.673pt}}
\multiput(1180.17,477.60)(30.000,-41.602){2}{\rule{0.400pt}{0.337pt}}
\multiput(1211.58,432.76)(0.497,-0.853){57}{\rule{0.120pt}{0.780pt}}
\multiput(1210.17,434.38)(30.000,-49.381){2}{\rule{0.400pt}{0.390pt}}
\multiput(1241.58,381.32)(0.497,-0.987){55}{\rule{0.120pt}{0.886pt}}
\multiput(1240.17,383.16)(29.000,-55.161){2}{\rule{0.400pt}{0.443pt}}
\multiput(1270.58,324.32)(0.497,-0.988){57}{\rule{0.120pt}{0.887pt}}
\multiput(1269.17,326.16)(30.000,-57.160){2}{\rule{0.400pt}{0.443pt}}
\multiput(1300.58,265.43)(0.497,-0.954){57}{\rule{0.120pt}{0.860pt}}
\multiput(1299.17,267.22)(30.000,-55.215){2}{\rule{0.400pt}{0.430pt}}
\multiput(1330.58,208.78)(0.497,-0.848){55}{\rule{0.120pt}{0.776pt}}
\multiput(1329.17,210.39)(29.000,-47.390){2}{\rule{0.400pt}{0.388pt}}
\multiput(1359.58,160.59)(0.497,-0.600){57}{\rule{0.120pt}{0.580pt}}
\multiput(1358.17,161.80)(30.000,-34.796){2}{\rule{0.400pt}{0.290pt}}
\multiput(1389.00,125.92)(1.084,-0.494){25}{\rule{0.957pt}{0.119pt}}
\multiput(1389.00,126.17)(28.013,-14.000){2}{\rule{0.479pt}{0.400pt}}
\put(262.0,113.0){\rule[-0.200pt]{6.986pt}{0.400pt}}
\end{picture}
 
\setlength{\unitlength}{0.240900pt}
\ifx\plotpoint\undefined\newsavebox{\plotpoint}\fi
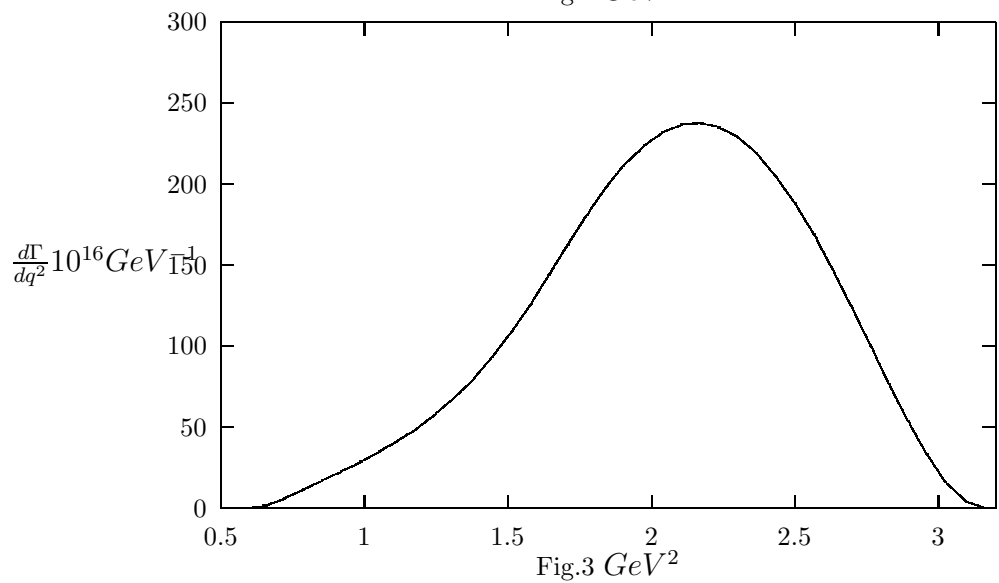
\begin{picture}(1500,900)(0,0)
\font\gnuplot=cmr10 at 10pt
\gnuplot
\sbox{\plotpoint}{\rule[-0.200pt]{0.400pt}{0.400pt}}%
\put(220.0,113.0){\rule[-0.200pt]{292.934pt}{0.400pt}}
\put(220.0,113.0){\rule[-0.200pt]{4.818pt}{0.400pt}}
\put(198,113){\makebox(0,0)[r]{0}}
\put(1416.0,113.0){\rule[-0.200pt]{4.818pt}{0.400pt}}
\put(220.0,240.0){\rule[-0.200pt]{4.818pt}{0.400pt}}
\put(198,240){\makebox(0,0)[r]{50}}
\put(1416.0,240.0){\rule[-0.200pt]{4.818pt}{0.400pt}}
\put(220.0,368.0){\rule[-0.200pt]{4.818pt}{0.400pt}}
\put(198,368){\makebox(0,0)[r]{100}}
\put(1416.0,368.0){\rule[-0.200pt]{4.818pt}{0.400pt}}
\put(220.0,495.0){\rule[-0.200pt]{4.818pt}{0.400pt}}
\put(198,495){\makebox(0,0)[r]{150}}
\put(1416.0,495.0){\rule[-0.200pt]{4.818pt}{0.400pt}}
\put(220.0,622.0){\rule[-0.200pt]{4.818pt}{0.400pt}}
\put(198,622){\makebox(0,0)[r]{200}}
\put(1416.0,622.0){\rule[-0.200pt]{4.818pt}{0.400pt}}
\put(220.0,750.0){\rule[-0.200pt]{4.818pt}{0.400pt}}
\put(198,750){\makebox(0,0)[r]{250}}
\put(1416.0,750.0){\rule[-0.200pt]{4.818pt}{0.400pt}}
\put(220.0,877.0){\rule[-0.200pt]{4.818pt}{0.400pt}}
\put(198,877){\makebox(0,0)[r]{300}}
\put(1416.0,877.0){\rule[-0.200pt]{4.818pt}{0.400pt}}
\put(220.0,113.0){\rule[-0.200pt]{0.400pt}{4.818pt}}
\put(220,68){\makebox(0,0){0.5}}
\put(220.0,857.0){\rule[-0.200pt]{0.400pt}{4.818pt}}
\put(445.0,113.0){\rule[-0.200pt]{0.400pt}{4.818pt}}
\put(445,68){\makebox(0,0){1}}
\put(445.0,857.0){\rule[-0.200pt]{0.400pt}{4.818pt}}
\put(670.0,113.0){\rule[-0.200pt]{0.400pt}{4.818pt}}
\put(670,68){\makebox(0,0){1.5}}
\put(670.0,857.0){\rule[-0.200pt]{0.400pt}{4.818pt}}
\put(896.0,113.0){\rule[-0.200pt]{0.400pt}{4.818pt}}
\put(896,68){\makebox(0,0){2}}
\put(896.0,857.0){\rule[-0.200pt]{0.400pt}{4.818pt}}
\put(1121.0,113.0){\rule[-0.200pt]{0.400pt}{4.818pt}}
\put(1121,68){\makebox(0,0){2.5}}
\put(1121.0,857.0){\rule[-0.200pt]{0.400pt}{4.818pt}}
\put(1346.0,113.0){\rule[-0.200pt]{0.400pt}{4.818pt}}
\put(1346,68){\makebox(0,0){3}}
\put(1346.0,857.0){\rule[-0.200pt]{0.400pt}{4.818pt}}
\put(220.0,113.0){\rule[-0.200pt]{292.934pt}{0.400pt}}
\put(1436.0,113.0){\rule[-0.200pt]{0.400pt}{184.048pt}}
\put(220.0,877.0){\rule[-0.200pt]{292.934pt}{0.400pt}}
\put(45,495){\makebox(0,0){${d\Gamma\over dq^{2}}10^{16}GeV^{-1}$ }}
\put(828,23){\makebox(0,0){Fig.3   $GeV^{2}$}}
\put(220.0,113.0){\rule[-0.200pt]{0.400pt}{184.048pt}}
\put(262,113){\usebox{\plotpoint}}
\multiput(262.00,113.60)(4.137,0.468){5}{\rule{3.000pt}{0.113pt}}
\multiput(262.00,112.17)(22.773,4.000){2}{\rule{1.500pt}{0.400pt}}
\multiput(291.00,117.58)(1.272,0.492){21}{\rule{1.100pt}{0.119pt}}
\multiput(291.00,116.17)(27.717,12.000){2}{\rule{0.550pt}{0.400pt}}
\multiput(321.00,129.58)(1.084,0.494){25}{\rule{0.957pt}{0.119pt}}
\multiput(321.00,128.17)(28.013,14.000){2}{\rule{0.479pt}{0.400pt}}
\multiput(351.00,143.58)(1.048,0.494){25}{\rule{0.929pt}{0.119pt}}
\multiput(351.00,142.17)(27.073,14.000){2}{\rule{0.464pt}{0.400pt}}
\multiput(380.00,157.58)(1.084,0.494){25}{\rule{0.957pt}{0.119pt}}
\multiput(380.00,156.17)(28.013,14.000){2}{\rule{0.479pt}{0.400pt}}
\multiput(410.00,171.58)(1.010,0.494){27}{\rule{0.900pt}{0.119pt}}
\multiput(410.00,170.17)(28.132,15.000){2}{\rule{0.450pt}{0.400pt}}
\multiput(440.00,186.58)(0.913,0.494){29}{\rule{0.825pt}{0.119pt}}
\multiput(440.00,185.17)(27.288,16.000){2}{\rule{0.413pt}{0.400pt}}
\multiput(469.00,202.58)(0.838,0.495){33}{\rule{0.767pt}{0.119pt}}
\multiput(469.00,201.17)(28.409,18.000){2}{\rule{0.383pt}{0.400pt}}
\multiput(499.00,220.58)(0.793,0.495){35}{\rule{0.732pt}{0.119pt}}
\multiput(499.00,219.17)(28.482,19.000){2}{\rule{0.366pt}{0.400pt}}
\multiput(529.00,239.58)(0.631,0.496){43}{\rule{0.604pt}{0.120pt}}
\multiput(529.00,238.17)(27.746,23.000){2}{\rule{0.302pt}{0.400pt}}
\multiput(558.00,262.58)(0.576,0.497){49}{\rule{0.562pt}{0.120pt}}
\multiput(558.00,261.17)(28.834,26.000){2}{\rule{0.281pt}{0.400pt}}
\multiput(588.00,288.58)(0.516,0.497){55}{\rule{0.514pt}{0.120pt}}
\multiput(588.00,287.17)(28.934,29.000){2}{\rule{0.257pt}{0.400pt}}
\multiput(618.58,317.00)(0.497,0.586){55}{\rule{0.120pt}{0.569pt}}
\multiput(617.17,317.00)(29.000,32.819){2}{\rule{0.400pt}{0.284pt}}
\multiput(647.58,351.00)(0.497,0.650){57}{\rule{0.120pt}{0.620pt}}
\multiput(646.17,351.00)(30.000,37.713){2}{\rule{0.400pt}{0.310pt}}
\multiput(677.58,390.00)(0.497,0.718){57}{\rule{0.120pt}{0.673pt}}
\multiput(676.17,390.00)(30.000,41.602){2}{\rule{0.400pt}{0.337pt}}
\multiput(707.58,433.00)(0.497,0.813){55}{\rule{0.120pt}{0.748pt}}
\multiput(706.17,433.00)(29.000,45.447){2}{\rule{0.400pt}{0.374pt}}
\multiput(736.58,480.00)(0.497,0.802){57}{\rule{0.120pt}{0.740pt}}
\multiput(735.17,480.00)(30.000,46.464){2}{\rule{0.400pt}{0.370pt}}
\multiput(766.58,528.00)(0.497,0.785){57}{\rule{0.120pt}{0.727pt}}
\multiput(765.17,528.00)(30.000,45.492){2}{\rule{0.400pt}{0.363pt}}
\multiput(796.58,575.00)(0.497,0.726){55}{\rule{0.120pt}{0.679pt}}
\multiput(795.17,575.00)(29.000,40.590){2}{\rule{0.400pt}{0.340pt}}
\multiput(825.58,617.00)(0.497,0.617){57}{\rule{0.120pt}{0.593pt}}
\multiput(824.17,617.00)(30.000,35.769){2}{\rule{0.400pt}{0.297pt}}
\multiput(855.00,654.58)(0.516,0.497){55}{\rule{0.514pt}{0.120pt}}
\multiput(855.00,653.17)(28.934,29.000){2}{\rule{0.257pt}{0.400pt}}
\multiput(885.00,683.58)(0.692,0.496){39}{\rule{0.652pt}{0.119pt}}
\multiput(885.00,682.17)(27.646,21.000){2}{\rule{0.326pt}{0.400pt}}
\multiput(914.00,704.58)(1.272,0.492){21}{\rule{1.100pt}{0.119pt}}
\multiput(914.00,703.17)(27.717,12.000){2}{\rule{0.550pt}{0.400pt}}
\put(944,716.17){\rule{6.100pt}{0.400pt}}
\multiput(944.00,715.17)(17.339,2.000){2}{\rule{3.050pt}{0.400pt}}
\multiput(974.00,716.93)(2.171,-0.485){11}{\rule{1.757pt}{0.117pt}}
\multiput(974.00,717.17)(25.353,-7.000){2}{\rule{0.879pt}{0.400pt}}
\multiput(1003.00,709.92)(0.945,-0.494){29}{\rule{0.850pt}{0.119pt}}
\multiput(1003.00,710.17)(28.236,-16.000){2}{\rule{0.425pt}{0.400pt}}
\multiput(1033.00,693.92)(0.576,-0.497){49}{\rule{0.562pt}{0.120pt}}
\multiput(1033.00,694.17)(28.834,-26.000){2}{\rule{0.281pt}{0.400pt}}
\multiput(1063.58,666.58)(0.497,-0.603){55}{\rule{0.120pt}{0.583pt}}
\multiput(1062.17,667.79)(29.000,-33.790){2}{\rule{0.400pt}{0.291pt}}
\multiput(1092.58,631.26)(0.497,-0.701){57}{\rule{0.120pt}{0.660pt}}
\multiput(1091.17,632.63)(30.000,-40.630){2}{\rule{0.400pt}{0.330pt}}
\multiput(1122.58,588.87)(0.497,-0.819){57}{\rule{0.120pt}{0.753pt}}
\multiput(1121.17,590.44)(30.000,-47.436){2}{\rule{0.400pt}{0.377pt}}
\multiput(1152.58,539.44)(0.497,-0.953){55}{\rule{0.120pt}{0.859pt}}
\multiput(1151.17,541.22)(29.000,-53.218){2}{\rule{0.400pt}{0.429pt}}
\multiput(1181.58,484.32)(0.497,-0.988){57}{\rule{0.120pt}{0.887pt}}
\multiput(1180.17,486.16)(30.000,-57.160){2}{\rule{0.400pt}{0.443pt}}
\multiput(1211.58,425.21)(0.497,-1.022){57}{\rule{0.120pt}{0.913pt}}
\multiput(1210.17,427.10)(30.000,-59.104){2}{\rule{0.400pt}{0.457pt}}
\multiput(1241.58,364.09)(0.497,-1.057){55}{\rule{0.120pt}{0.941pt}}
\multiput(1240.17,366.05)(29.000,-59.046){2}{\rule{0.400pt}{0.471pt}}
\multiput(1270.58,303.37)(0.497,-0.971){57}{\rule{0.120pt}{0.873pt}}
\multiput(1269.17,305.19)(30.000,-56.187){2}{\rule{0.400pt}{0.437pt}}
\multiput(1300.58,245.65)(0.497,-0.887){57}{\rule{0.120pt}{0.807pt}}
\multiput(1299.17,247.33)(30.000,-51.326){2}{\rule{0.400pt}{0.403pt}}
\multiput(1330.58,193.12)(0.497,-0.743){55}{\rule{0.120pt}{0.693pt}}
\multiput(1329.17,194.56)(29.000,-41.561){2}{\rule{0.400pt}{0.347pt}}
\multiput(1359.00,151.92)(0.516,-0.497){55}{\rule{0.514pt}{0.120pt}}
\multiput(1359.00,152.17)(28.934,-29.000){2}{\rule{0.257pt}{0.400pt}}
\multiput(1389.00,122.92)(1.392,-0.492){19}{\rule{1.191pt}{0.118pt}}
\multiput(1389.00,123.17)(27.528,-11.000){2}{\rule{0.595pt}{0.400pt}}
\end{picture}
 
In conclusion, the branching ratios and distribution functions of the four
decay modes of $\tau\rightarrow\rho\pi\pi\nu$ have been calculated in terms
of the effective chiral theory of mesons[1]. The study is parameter free and
theoretical results agree with the data. The theory predicts the $a_{1}$
dominance in these four decay modes.
 
The author thanks discussion with K.K.Gan and J.Smith.
This research was partially
supported by DOE Grant No. DE-91ER75661.
 
{\large Appendix}\\
Contributions of the diagrams to the matrix 
element(18):
\begin{enumerate}
\item diagrams involving the vertices(13-15)
\begin{eqnarray}
\lefteqn{f_{12}=-\frac{1}{\pi^{2}g^{2}f^{2}_{\pi}}\{{4c\over g}(1-{c\over g})
+{1\over \pi^{2}g^{2}}(1-{2c\over g})^{2}+2(1-{2c\over g})[1+\frac{m^{2}_{\rho}}
{2\pi^{2}f^{2}_{\pi}}[(1-{2c\over g})^{2}-4\pi^{2}c^{2}]\}}\nonumber \\
&&\{4(1-{c\over g})+{2q^{2}\over\pi^{2}f^{2}_{\pi}}(1-{2c\over g})^{2}
+{2k^{2}_{1}\over\pi^{2}f^{2}_{\pi}}(1-{c\over g})(1-{2c\over g})\}
-{4k^{2}_{1}\over\pi^{4}g^{2}f^{4}_{\pi}}
(1-{c\over g})^{2}(1-{2c\over g}).
\end{eqnarray}
\item diagrams(fig.1(c))
\begin{eqnarray}
\lefteqn{f_{12}={1\over k^{2}_{1}}{16\over g^{2}}\{1+{q^{2}\over2\pi^{2}
f^{2}_{\pi}}[(1-{2c\over g})^{2}-4\pi^{2}c^{2}]+{k^{2}_{1}\over2\pi^{2}
f^{2}_{\pi}}(1-{c\over g})(1-{2c\over g})\}}\nonumber \\
&&\{1+\frac{m^{2}_{\rho}}{2\pi^{2}f^{2}_{\pi}}[(1-{2c\over g})^{2}
-4\pi^{2}c^{2}]\}+{8\over \pi^{2}g^{2}f^{2}_{\pi}}(1-{c\over g})
(1-{2c\over g}).
\end{eqnarray}
\item diagrams(fig.1(a))
\begin{eqnarray}
\lefteqn{f={4\over f^{2}_{\pi}}\{{F^{2}\over g^{2}}+(q^{2}+m^{2}_{\rho})
[{2c^{2}\over g^{2}}-{1\over4\pi^{2}g^{2}}(1-{2c\over g})^{2}+
{3\over4\pi^{2}g^{2}}(1-{2c\over g})]}\nonumber \\
&&+k^{2}_{1}
[-{2c^{2}\over g^{2}}+{1\over4\pi^{2}g^{2}}(1-{2c\over g})^{2}-{3\over2\pi^{2}g^{2}}
(1-{2c\over g})+{1\over2\pi^{2}g^{2}}]+k^{2}_{2}[-{2c^{2}\over g^{2}}+
{1\over4\pi^{2}g^{2}}(1-{2c\over g})^{2}]\},\nonumber \\
&&f_{11}=-{2\over\pi^{2}g^{2}f^{2}_{\pi}}(1-{2c\over g}),\nonumber \\
&&f_{12}=-{8\over f^{2}_{\pi}}\{-{2c^{2}\over g^{2}}+{3\over4\pi^{2}g^{2}}
(1-{2c\over g})^{2}+{3\over2\pi^{2}g^{2}}(1-{2c\over g})\},\nonumber \\
&&f_{21}={8\over f^{2}_{\pi}}\{-{4c^{2}\over g^{2}}+{3\over4\pi^{2}g^{2}}
(1-{2c\over g})\},\nonumber \\
&&f_{22}=f_{11}.
\end{eqnarray}
\item diagrams(fig.1(d))
\begin{eqnarray}
\lefteqn{BWR=\frac{1}{k^{2}_{3}-m^{2}_{\rho}
+i\sqrt{k^{2}_{3}}\Gamma_{\rho}(k^{2}_{3})}},\nonumber \\
&&f={4\over g^{2}}BWR(k^{2}_{1}-k^{2}_{2})\{1+
\frac{p_{1}\cdot p_{2}}{\pi^{2}f^{2}_{\pi}}[(1-{2c\over g})^{2}
-4\pi^{2}c^{2}]\},\nonumber \\
&&f_{12}={16\over g^{2}}BWR\{1+
\frac{p_{1}\cdot p_{2}}{\pi^{2}f^{2}_{\pi}}[(1-{2c\over g})^{2}
-4\pi^{2}c^{2}]\},\nonumber \\
&&f_{21}=-f_{12},
\end{eqnarray}
where \(k_{3}=q-p\).
\end{enumerate}

\newpage
\leftline{\bf Figure Captions}
\begin{description}
\item[Fih. 1] Diagrams of the decay $\tau\rightarrow\rho\pi\pi\nu$
\item[Fig. 2] Distribution function of $\tau\rightarrow\rho^{0}\pi^{0}\pi^{-}
(\rho^{-}\pi^{-}\pi^{+})\nu$
\item[Fig. 3] Distribution function of $\tau\rightarrow
\rho^{+}\pi^{-}\pi^{-}
(\rho^{-}\pi^{0}\pi^{0})\nu$

\end{description} 
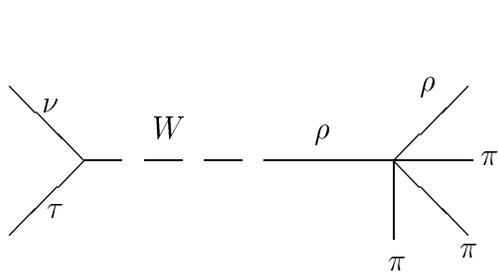
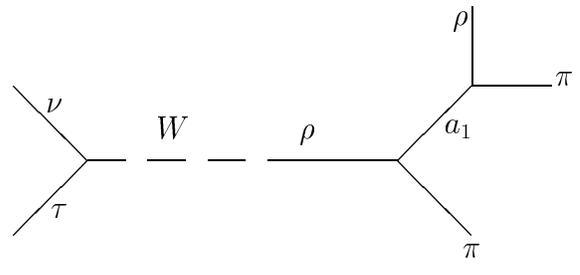
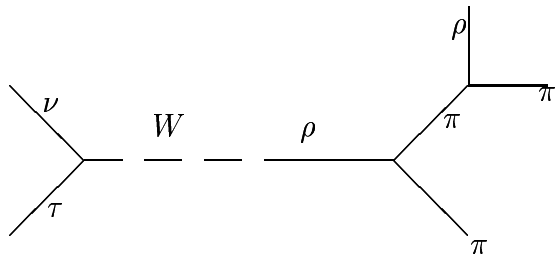
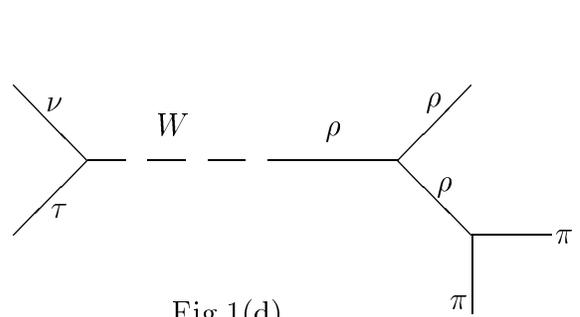
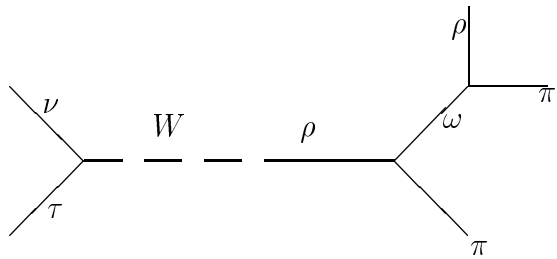
\end{document}